\documentclass[letterpaper, 10 pt, conference]{ieeeconf}
\usepackage[utf8]{inputenc}

\IEEEoverridecommandlockouts

\usepackage{balance}

\usepackage{multirow}

\usepackage{amsmath,mathtools} 
\usepackage{amsfonts,amssymb}
\usepackage{bbm}

\usepackage{xspace}

\usepackage{graphicx}
\graphicspath{{figs/}}

\usepackage{subcaption}
\usepackage{caption}

\usepackage{url}
\usepackage{hyperref}
\hypersetup{colorlinks=true, linkcolor= blue, allcolors=blue, citecolor = blue, filecolor=black, urlcolor=blue, breaklinks=true}

\usepackage{todonotes}
\usepackage{comment}

\usepackage{tikz}
\usetikzlibrary{arrows,positioning}  

\usepackage{pgfplots}\pgfplotsset{compat=1.9}
\usepackage{grffile}
\pgfplotsset{compat=newest} 
\usetikzlibrary{plotmarks}
\usetikzlibrary{arrows.meta}
\usepgfplotslibrary{patchplots}
\usepgfplotslibrary{groupplots}

\pgfplotsset{compat=1.15}
\usepackage{mathrsfs}
\usetikzlibrary{arrows}

\usepackage{tabularx}

\usepackage[ruled]{algorithm}
\usepackage[noend]{algpseudocode}

\usepackage{diagbox}
\newcolumntype{K}[1]{>{\centering\arraybackslash}p{#1}}

\makeatletter
\newcommand{\multiline}[1]{
  \begin{tabularx}{\dimexpr\linewidth-\ALG@thistlm}[t]{@{}X@{}}
    #1
  \end{tabularx}
}
\makeatother

\usepackage[per-mode=symbol]{siunitx}

\usepackage{cite} 
\usepackage{array} 
\usepackage{booktabs}

\newcounter{remark}
\newenvironment{remark}{
\par\vspace{3pt}\noindent\refstepcounter{remark}\textbf{Remark~\theremark:}}
{\par\endtrivlist\unskip}

\newcounter{problem}

\usepackage[extramath,probability,reviewmark]{mylatexdefs}

\newcommand{\GPModel}{\ensuremath{\GGG \PPP}}

\newcommand{\GPmean}{\ensuremath{\mu}}

\newcommand{\GPstdvar}{\ensuremath{\sigma}}

\newcommand{\bb}[1]{\ensuremath{\mathbf{#1}}}
\newcommand{\bbsym}[1]{\ensuremath{\boldsymbol{#1}}}
\newcommand{\CAV}[1]{CAV\textendash \ensuremath{#1}\xspace}
\newcommand{\HDV}[1]{HDV\textendash\ensuremath{#1}\xspace}
\newcommand{\vehicle}[1]{vehicle\textendash\ensuremath{#1}\xspace}

\usepackage{stfloats}

\def\BibTeX{{\rm B\kern-.05em{\sc i\kern-.025em b}\kern-.08em
    T\kern-.1667em\lower.7ex\hbox{E}\kern-.125emX}}
\begin{document}

\title{\LARGE \bf Controller Adaptation via Learning Solutions of\\Contextual Bayesian Optimization}
\author{Viet-Anh Le$^{1,2}$, {\IEEEmembership{Student Member, IEEE}}, and Andreas A. Malikopoulos$^{3}$, {\IEEEmembership{Senior Member, IEEE}}
\thanks{This research was supported in part by NSF under Grants CNS-2401007, CMMI-2348381, IIS-2415478, and in part by MathWorks.}
\thanks{$^{1}$Department of Mechanical Engineering, University of Delaware, Newark, DE 19716 USA.}
\thanks{$^{2}$Systems Engineering Field, Cornell University, Ithaca, NY 14850 USA.}
\thanks{$^{3}$School of Civil and Environmental Engineering, Cornell University, Ithaca, NY 14853 USA.}
\thanks{Emails: {\tt\small \{vl299,amaliko\}@cornell.edu}.}
}

\maketitle

\begin{abstract}
In this work, we propose a framework for adapting the controller's parameters based on learning optimal solutions from contextual black-box optimization problems.
We consider a class of control design problems for dynamical systems operating in different environments or conditions represented by contextual parameters. 
The overarching goal is to identify the controller parameters that maximize the controlled system's performance, given different realizations of the contextual parameters.
We formulate a contextual Bayesian optimization problem in which the solution is actively learned using Gaussian processes to approximate the controller adaptation strategy. We demonstrate the efficacy of the proposed framework with a sim-to-real example. We learn the optimal weighting strategy of a model predictive control for connected and automated vehicles interacting with human-driven vehicles from simulations and then deploy it in a real-time experiment.
\end{abstract}

\section{Introduction}

Controller tuning generally aims to find controller parameters that optimize specific performance metrics.
In recent years, controller tuning has received increasing attention in different control applications.
Some popular approaches that have been presented in the literature include 
reinforcement learning
\cite{mehndiratta2018automated,zanon2020safe,kordabad2023reinforcement,romero2024actor},
differentiable programming \cite{amos2018differentiable,tao2024difftune},
Bayesian optimization (BO) \cite{neumann2019data,schillinger2017safe,muratore2021data,turchetta2020robust,edwards2021automatic},
self-learning control \cite{Malikopoulos2010a}, and Kalman filtering \cite{menner2023automated,Allamaa2022Sim2real}. 
While classical controller tuning approaches often focus on optimizing the controller for invariant systems, in practice, the system must operate under changing conditions, environments, or tasks, such as throttle valve systems with different goal positions \cite{bischoff2013learning}, 
legged robots walking on various surfaces \cite{yu2020learning},
advanced powertrain systems operating under different driving styles \cite{Malikopoulos2008b,Malikopoulos2011}, 
or connected and automated vehicles \cite{Malikopoulos2020} interacting with different styles of human-driven vehicles \cite{Le2023ACC}.
In such situations, the controllers may need to be adapted to account for the varying factors.
A potential approach to address this problem is \emph{contextual BO} \cite{krause2011contextual}, which is an extension of BO that considers additional variables beyond the optimization variables.
Berkenkamp \etal \cite{berkenkamp2023bayesian} presented a safe contextual BO framework in which the safe control parameters for the unobserved contexts can be found given the surrogate model transferred from observed contexts.
Fr\"{o}hlich \etal \cite{frohlich2022contextual} considered a learned dynamic model as contexts in contextual BO to transfer knowledge across different environmental conditions in autonomous racing. 
Xu \etal \cite{xu2024data} proposed primal-dual contextual BO to handle time-varying disturbances for thermal control systems of smart buildings.
Stenger \etal \cite{stenger2023vehicle} combined contextual BO with constrained max-value entropy search to optimize the MPC parameters for vehicle climate control, considering different passenger-defined mass flow levels.

In this letter, we propose a framework that approximates a \emph{controller adaptation strategy} for dynamical systems by leveraging the optimal solutions of contextual BO.
We formulate the controller adaptation problem as a contextual BO problem in which the varying system and controller parameters are treated as contexts and optimization variables, respectively.
We employ Gaussian processes (GPs) to learn the latent mapping from the contexts to the solutions of the BO problem and utilize an adaptive sampling technique for the contexts.
Therefore, our proposed framework fits well in applications where the contexts can be sampled, such as \emph{sim-to-real applications} or \emph{optimal experiment design problems}. 
The strategy learned from data obtained in observed situations can be utilized in unobserved situations to facilitate real-time adaptation of the control parameters.
Our work differs from the related work \cite{berkenkamp2023bayesian,frohlich2022contextual,xu2024data,stenger2023vehicle} in the following aspects.
First, in \cite{frohlich2022contextual,xu2024data,stenger2023vehicle}, the contexts were set by the environment, while our work considers the problem where the context can be sampled.
In \cite{berkenkamp2023bayesian}, the contexts are selected manually, whereas we decide the contexts to sample to efficiently approximate the latent mapping of the solutions. 
The concept of approximating the latent mapping from contexts to solutions in contextual BO was previously explored in \cite{stenger2023vehicle} and \cite[Appendix~D]{char2019offline}.
In \cite{stenger2023vehicle}, the authors applied contextual BO for a discretized context set and used interpolation to approximate the solutions, whereas our framework actively samples contexts from a continuous context space.
In \cite[Appendix~D]{char2019offline}, the contexts are sampled based on the surrogate model, and the solution for a new context is determined by optimizing the posterior mean of the final surrogate model.
Meanwhile, we propose approximating the solution model by a GP combined with an outer-loop adaptive sampling algorithm for selecting the contexts so that we do not need to solve an optimization problem to find solutions for new contexts in real time.

We demonstrate the effectiveness of the framework in a sim-to-real example related to model predictive control (MPC) for connected and automated vehicles (CAVs) interacting with human-driven vehicles (HDVs).
In this application, the objective weights, which characterize human driving behavior, are treated as contexts, while the objective weights for the CAV are optimization variables. 
Since the contexts may vary over time in real-time deployment, the controller must quickly and effectively adapt to accommodate diverse and time-varying contexts.
Using our framework, we provide a weight adaptation strategy for the MPC given different HDV driving behaviors from simulations so that the desired performance can be achieved.
We perform real-time experiments, in which the learned strategy is then utilized alongside the real human driving behavior obtained from inverse reinforcement learning (IRL) \cite{kuderer2015learning} to adapt the MPC.
Our code for the implementation of the proposed framework, the MPC weight adaptation example, and other examples is available at \url{https://github.com/vietanhle0101/Learn-Contextual-BayesOpt}.

The remainder of the letter is organized as follows.
In Section~\ref{sec:problem}, we provide the problem statement for controller adaptation.
We present the framework for learning the controller adaptation strategy in Section~\ref{sec:bo}.
We demonstrate the framework and show the results in Section~\ref{sec:example}.
Finally, we draw some conclusions in Section~\ref{sec:concl}.

\section{Controller Adaptation Problem} \label{sec:problem}

We consider the problem of designing a controller for a dynamic system with contextual parameters, which can vary depending on the tasks or change over time due to environmental conditions.  
For instance, these contextual parameters could represent the setpoints at which the system operates, the weights in a cost function describing system behavior, or the time-varying coefficients of the system dynamics.
Let $\bbsym{\theta} \in \bbsym{\Theta}$ be a vector representing the system contextual parameters with a set of values \bbsym{\Theta}.
We consider the system controlled by a controller with parameters that can be tuned or adapted to optimize performance.
For example, the controller parameters might encompass the gains of a PID controller, the coefficients of a state-feedback control law, or the weights within an MPC cost function. 
Let ${\bbsym{z} \in \ZZZ}$ be the vector of controller parameters.
In the controller tuning problem, if the system contextual parameters $\bbsym{\theta}$ are fixed, we seek the optimal controller parameters $\bbsym{z}^*$ so that a certain performance metric is optimized, \ie
\begin{equation}
\label{eq:main_prob}
\underset{\bbsym{z} \in \ZZZ} \maximize \quad \;
J (\bbsym{z}, \bbsym{\theta}),
\end{equation}
where $J : \ZZZ \times \Theta \rightarrow \RR$ is a performance metric function.
We are interested in the problem of finding an adaptation strategy $\bbsym{\gamma}$ for control parameters $\bbsym{z}$ given different realizations of $\bbsym{\theta}$ in \eqref{eq:main_prob}.
The objective of the proposed framework is to learn the latent function $\bbsym{\gamma} : \Theta \rightarrow \ZZZ$ in \eqref{eq:main_prob}, \ie
\begin{equation}
\bbsym{z}^* = \bbsym{\gamma} (\bbsym{\theta}).
\end{equation}
This solution can be valuable in real-time control applications, where using controller tuning is rather infeasible,  as it can be used to adapt $\bbsym{z}$ given the values of $\bbsym{\theta}$ in case some system parameters have changed. 
In our approach, we consider that the performance metric $J$ has the following properties:
\begin{itemize}
\item $J$ is a black box of $\bbsym{z}$ and $\bbsym{\theta}$, \ie we do not have an analytical expression for $J$ in terms of $\bbsym{z}$ and $\bbsym{\theta}$. 
\item We can only observe the output of $J$ by evaluating the state and input trajectories of the system through simulations or experiments; however, we do not have access to the first- or second-order derivatives.
\item Observations of $J$ can be noisy with independent and identically distributed (i.i.d.) Gaussian noise.
\item Obtaining the observations of $J$ may be expensive and/or complex. For example, it may involve conducting experiments or requiring mass simulations with different initial conditions to obtain the average performance metric.
\end{itemize}

Given the above properties of the performance metric and if the system contextual parameters $\bbsym{\theta}$ are fixed, BO \cite{jones1998efficient} is a commonly used method to tune the controller parameters.
In contrast, for systems with varying parameters, one can utilize contextual BO \cite{krause2011contextual}, an extension of BO that considers additional variables known as contexts.
Therefore, in the next section, we recast \eqref{eq:main_prob} as a contextual black-box optimization problem in which $\bbsym{z}$ and $\bbsym{\theta}$ are considered as the optimization variable and the context, respectively, and propose to approximate the solutions of contextual BO using GPs.
Note that though we considered an unconstrained optimization problem in \eqref{eq:main_prob}, black-box constraints can be taken into account by using penalty functions.
An alternative approach is to use GPs to learn black-box constraints and formulate probabilistic constraints, \eg \cite{stenger2023vehicle}.

\section{Learning Solutions of Contextual Black-box Optimization} \label{sec:bo}

In this section, we first provide background information on GPs and contextual BO, followed by a discussion of the method for approximating contextual BO solutions.

\subsection{Single-Output and Multi-Output Gaussian Processes}

A GP defines a distribution over functions where any finite subset of function values follows a multivariate Gaussian distribution \cite{liu2018gaussian}.
A GP model of a scalar function $f (\bbsym{x})$, denoted as $\GPModel_f (\bbsym{x})$, is specified by a mean function $m(\bbsym{x})$ and a covariance function (kernel) $\kappa(\bbsym{x},\bbsym{x}')$ which are parameterized by some hyperparameters.
Given a training dataset $\DDD = (\bbsym{X}, \bbsym{Y})$, where $\bbsym{X} = [\bbsym{x}_1^\top, \dots, \bbsym{x}_N^\top]^\top$ and $\bbsym{Y} = [y_1, \dots, y_N]^\top$ are concatenated vectors of $N\in\mathbb{N}$ observed inputs and corresponding outputs, those hyperparameters can be learned by maximizing the likelihood.
Without loss of generality, we consider a zero-mean function in our exposition.
At a new input $\bbsym{x}_*$, the GP prediction is a Gaussian distribution $\NNN (\mu_*, \sigma_*)$ that is computed by
\begin{subequations}
\label{eq:gp-regression}
\begin{align}
\mu_* &= \bb{K}_* (\bb{K}+\sigma_n^2 \II)^{-1} \bb{Y}, \label{eq:gpmean}\\
\sigma_*^2 & = \bb{K}_{**} - \bb{K}_* (\bb{K}+\sigma_n^2\, \II_N)^{-1} \bb{K}_*^\top , \label{eq:gpvar}
\end{align}
\end{subequations}
where \(K_* = [\kappa(\bbsym{x}_*, \bbsym{x}_1), \dots, \kappa(\bbsym{x}_*, \bbsym{x}_N)]\), \(K_{* *} = \kappa(\bbsym{x}_*, \bbsym{x}_*)\), $K$ is the covariance matrix with elements \(K_{ij} = \kappa(\bbsym{x}_i, \bbsym{x}_j)\), $\sigma_n^2$ is the noise variance, and $\II_N$ is the $N \times N$ identity matrix.

GP can be extended to learn a multi-output function $\bbsym{f} (\bbsym{x}) = [f_1(\bbsym{x}), \dots, f_Q(\bbsym{x})]$ with $Q \in \NN_{>1}$ outputs.
A multi-output GP can be specified by a multi-output kernel ${\KKK(\bbsym{x}, \bbsym{x}') = \left[ \kappa_{ij} (\bbsym{x}, \bbsym{x}') \right]} \in \RR^{Q \times Q}$, for $i,j = 1, \dots, Q$.
There are several approaches for multi-output kernels (see \cite[Chapter 4]{alvarez2012kernels} for a review of commonly used approaches).
In this paper, we consider the intrinsic coregionalization model (ICM) \cite{bonilla2007multi}, which yields the following multi-output kernel, 
\begin{equation}
\KKK (\bbsym{x}, \bbsym{x}') = \kappa(\bbsym{x}, \bbsym{x}') \bb{B},
\end{equation}
where $\bb{B} \in \RR^{Q \times Q}$ is a symmetric and positive semidefinite matrix describing the correlation between different outputs.
The prediction of a multi-output GP includes the following predictive mean vector and covariance matrix
\begin{subequations}
\label{eq:mogp}
\begin{align}
\bbsym{\mu}_* &= \bb{K}^{(M)}_* (\bb{K}^{(M)}+\sigma_n^2 \II)^{-1} \bb{Y}, \label{eq:mogpmean}\\
\bbsym{\Sigma}_* &= \bb{K}^{(M)}_{* *} - \bb{K}^{(M)}_* (\bb{K}^{(M)}+\sigma_n^2\, \II_{NQ} )^{-1} (\bb{K}^{(M)}_*)^\top , \label{eq:mogpvar}
\end{align}
\end{subequations}  
where $\bb{K}^{(M)} = \bb{B}\otimes \bb{K}$, 
$\bb{K}^{(M)}_* = \bb{B}\otimes \bb{K}_*$, 
$\bb{K}^{(M)}_{**} = \bb{B}\otimes \bb{K}_{**}$, in which $\otimes$ represents the Kronecker product.

\subsection{Contextual Bayesian Optimization}

Contextual BO \cite{krause2011contextual} is an extension of BO that aims to solve a class of black-box optimization problems with contexts that are not part of the optimization variables.
Recall that in the controller adaptation problem, we aim at maximizing a black-box function $J(\bbsym{z},\, \bbsym{\theta})$ in \eqref{eq:main_prob}.
We define $\GPModel_o (\bbsym{z}, \bbsym{\theta})$ as the surrogate model that learns $J(\bbsym{z},\, \bbsym{\theta})$.
The kernel of the surrogate model can be formed by considering a product kernel of the kernels over context and variable spaces as follows \cite{krause2011contextual}
\begin{equation}
\kappa \big( (\bbsym{z}, \bbsym{\theta}), (\bbsym{z}', \bbsym{\theta}') \big)
= \kappa_z (\bbsym{z}, \bbsym{z}') . \kappa_{\theta} (\bbsym{\theta}, \bbsym{\theta}'),
\end{equation}
which implies that two context-variable pairs are similar if the contexts are similar and the variables are similar.
Given a realization of the context $\bbsym{\theta}$, the contextual BO is identical to the original BO, and the algorithm works as follows. 
First, it optimizes an acquisition function $\xi$ to find the next candidate of the solution,
\begin{equation}
\bbsym{z}^{(j)} = \underset{\bbsym{z} \in \ZZZ} \argmax \; \xi \big( \GPmean_{o} (\bbsym{z}, \bbsym{\theta}), \GPstdvar_o (\bbsym{z}, \bbsym{\theta}) \big),
\end{equation}
where $\GPmean_o$ and $\GPstdvar_o$ are the posterior mean and standard deviation of $\GPModel_o$.
For example, the upper confidence bound (UCB) acquisition function was used in contextual BO in \cite{krause2011contextual} given by
\begin{equation}
\xi \big( \GPmean_{o} (\bbsym{z}, \bbsym{\theta}), \GPstdvar_o (\bbsym{z}, \bbsym{\theta}) \big)
= \GPmean_o (\bbsym{z}, \bbsym{\theta}) + \beta^{1/2} \GPstdvar_o (\bbsym{z}, \bbsym{\theta}).
\end{equation}
Next, the output of the performance metric at that sampling candidate is evaluated and added to the training dataset to retrain the surrogate model. 
This process is repeated until a maximum number of iterations is reached and the best-evaluated candidate is returned.
In this letter, we use GPs to learn the latent mapping from the context $\bbsym{\theta}$ to the solution $\bbsym{z}^{*}$ returned from contextual BO, \ie $\bbsym{z}^{*} = \bbsym{\gamma} (\bbsym{\theta}) \sim \GPModel_s (\bbsym{\theta})$.
We call $\GPModel_s (\bbsym{\theta})$ as the solution model.
Depending on $\bbsym{z}$ is a scalar or a vector, we learn $\GPModel_s$ by a single-output GP or a multi-output GP, respectively.
Note that the latent function of the solution model may be non-smooth. 
There are several variants of GPs for handling such cases, for example, non-stationary kernels \cite{paciorek2003nonstationary} or deep kernel learning \cite{wilson2016deep}.
In this paper, we utilize deep kernel learning approach presented in \cite{wilson2016deep}.
In deep kernel learning, instead of using a base kernel $ \kappa(\bbsym{x}_i, \bbsym{x}_j)$, we construct a deep learning-based kernel
$\kappa( \pi (\bbsym{x}_i), \pi (\bbsym{x}_j)),$
where $\pi (\bbsym{x})$ is a deep neural network.
As a result, deep kernel learning can leverage both the nonparametric flexibility of GP base kernels and the structural properties of deep neural networks for learning highly nonlinear or non-smooth functions.
The hyperparameters of the base kernel and the weights of the neural network can be jointly trained by maximizing the log marginal likelihood.

\subsection{Adaptive Sampling for the Solution Model}

To efficiently and rapidly learn the solution model, we adopt the concept from Bayesian experimental design, aiming to find the set of most informative sampling points by maximizing the information gain.
Since finding the maximizer of information gain is NP-hard, a commonly employed approach is to use a greedy adaptive sampling algorithm.
At each iteration, the greedy algorithm selects the sampling location that maximizes the conditional entropy, and the GP model is recursively updated with the data obtained from the new sampling location.
For a multi-output GP, where the uncertainty is represented by the covariance matrix, maximizing the conditional entropy is equivalent to maximizing the log determinant of the covariance matrix \cite{le2022multistep}. 
Thus, the adaptive sampling optimization problem for the solution model at each iteration $k$ to find the next sampling location of $\bbsym{\theta}$ is formulated as follows
\begin{equation}
\label{eq:sampling}
\bbsym{\theta}^{(k)} = \operatorname*{argmax}_{\bbsym{\theta} \in \Theta}\;
\log \det \; \bbsym{\Sigma}_{s} (\bbsym{\theta}).
\end{equation}

\begin{algorithm}[b]
  \caption{Inner-loop Bayesian optimization}
  \label{alg:bayesopt}
    \begin{algorithmic}[1]
    \Require $k_{\mathrm{max}} \in \NN \setminus \{0\}$
    \Procedure{BayesOpt}{$\bbsym{\theta}$, $\GPModel_o$}
    \State $\GPModel_o^{(0)} \gets \GPModel_o$
    \For {$k = 1, \dots, k_{\mathrm{max}}$}
    \State Find the next solution candidate \(\bbsym{z}^{(k)}\) by optimizing acquisition function given $\GPModel_o^{(k-1)}$.
    \State Obtain an observation of the performance metric $J^{(k)} = J \Big( \bbsym{z}^{(k)}, \bbsym{\theta} \Big)$.
    \State Add $(\bbsym{z}^{(k)}, \bbsym{\theta}, J^{(k)})$ to $\DDD_o$ (see Remark~\ref{rem:data}) and re-train the surrogate GP model to obtain $\GPModel_o^{(k)}$.
    \EndFor 
    \State \textbf{return} $\bbsym{z}^{*} = \underset{\bbsym{z}}{\argmax} \; \bbsym{\mu}_o^{(k_{\max})} (\bbsym{z}, \bbsym{\theta})$, $\GPModel_o^{(k_{\max})}$
    \EndProcedure
  \end{algorithmic}
\end{algorithm}

The algorithm is presented in Algorithm~\ref{alg:sampling} and is summarized as follows.
At each iteration $j \in \NN$, we first propose the next sampling location $\bbsym{\theta}^{(j)}$ by solving the adaptive sampling optimization problem given the current solution model $\GPModel_s^{(j-1)}$.
Then, we fix $\bbsym{\theta}^{(j)}$ and apply an inner-loop contextual BO (Algorithm~\ref{alg:bayesopt}) to find the next candidate of the solution $\bbsym{z}^{(j)}$.
Once the solution is obtained, we update the training dataset and retrain $\GPModel_s$. 
These steps are repeated until a maximum number $j_{\max}$ of iterations is reached. 

\begin{remark}
\label{rem:data}
In Algorithm~\ref{alg:bayesopt}, we reuse the GP surrogate model trained on data from previous contexts to enable knowledge transfer to a new context \cite{berkenkamp2023bayesian,frohlich2022contextual}, which makes the training data size of $\GPModel_o$ larger over the iterations.
Thus, if the training dataset exceeds the maximum size, a heuristic rule \cite{kabzan2019learning} can be used so that the old data is replaced by new data observation.
\end{remark}

\begin{algorithm}[tb]
  \caption{Outer-loop adaptive sampling}
  \label{alg:sampling}
  \begin{algorithmic}[1]
    \Require $j_{\mathrm{max}} \in \NN \setminus \{0\}$, $\GPModel_s^{(0)}$, $\GPModel_o^{(0)}$
    \For {$j = 1, \dots, j_{\max}$}
    \State Find next sampling location \(\bbsym{\theta}^{(j)}\) of the context by solving the adaptive sampling problem given $\GPModel_s^{(j-1)}$.
    \State $\bbsym{z}^{(j)*}, \GPModel_o^{(j)} \gets$ \Call{BayesOpt}{$\bbsym{\theta}^{(j)}, \GPModel_o^{(j-1)}$} (see Algorithm~\ref{alg:bayesopt})
    \State Add $(\bbsym{\theta}^{(j)}, \bbsym{z}^{(j)*})$ to $\DDD_s$ and re-train the solution GP model to obtain $\GPModel_s^{(j)}$.
    \EndFor 
    \State \textbf{return} $\GPModel_s^{(j_{\max})}$
  \end{algorithmic}
\end{algorithm}

\section{Illustrative Example} \label{sec:example}

\begin{figure}[!t]
\centering
\includegraphics[scale=0.4, bb = 80 120 600 530, clip=true]{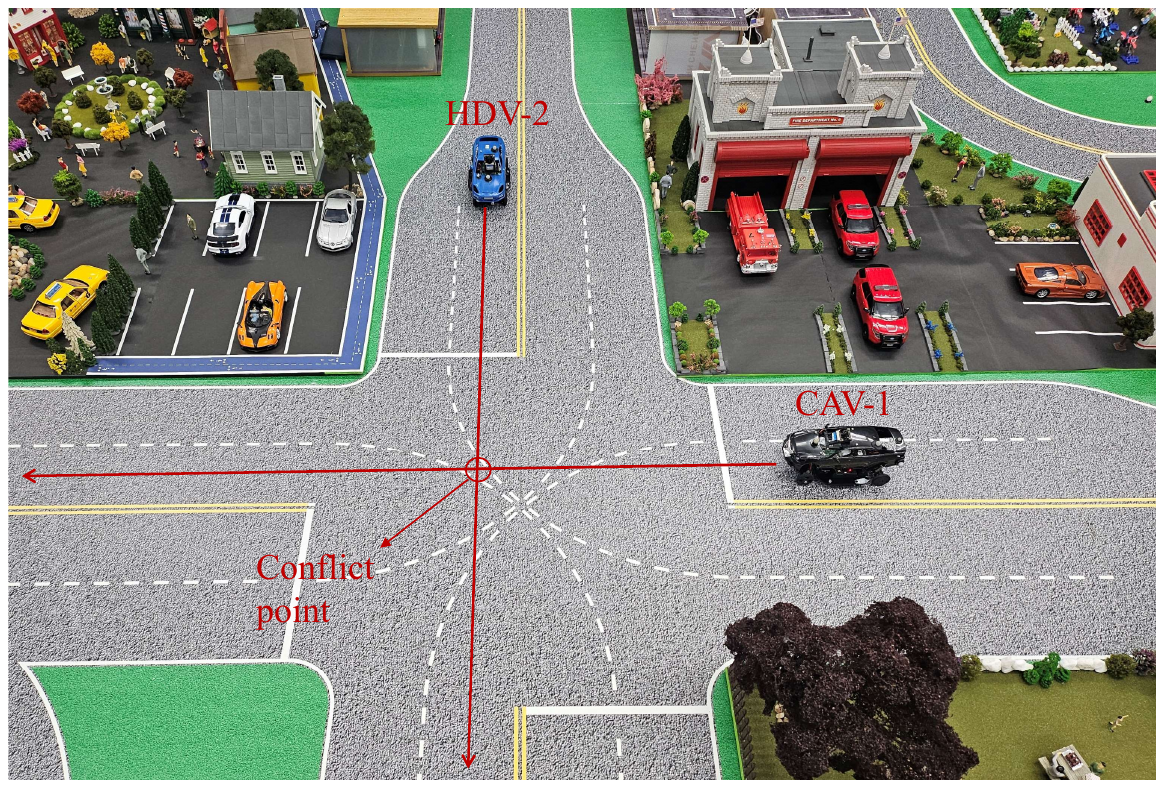}
\caption{An intersection scenario with a CAV and an HDV.}
\vspace{-3mm}
\label{fig:intersection}
\end{figure}

In this section, we demonstrate the proposed framework with an example of learning the weight adaptation strategy of MPC for CAVs while interacting with HDVs at a signal-free intersection.
We consider an intersection scenario in a robotic testbed called the Information and Decision Science Lab's Scaled Smart City (\texttt{IDS3C}) \cite{chalaki2021CSM} shown in Fig.~\ref{fig:intersection}.
We apply the proposed framework in a \emph{sim-to-real} manner, where the MPC weight adaptation strategy is learned from simulations and subsequently deployed to real-world experiments.

\subsection{Learning MPC Weight Adaptation Strategy for CAVs}
\label{subsec:mpc}

First, we summarize the potential game-based MPC formulation that we previously developed in \cite{Le2023ACC} for a CAV while interacting with an HDV.
Let \CAV{1} and \HDV{2} denote the vehicles involved in the intersection scenario.
The dynamics of each vehicle $i$ are given by the double-integrator dynamics
\begin{equation}
\label{eq:integrator}
\begin{split}
p_{i,k+1} &= p_{i,k} + \Delta T v_{i,k} + \frac{1}{2} \Delta T^2 a_{i,k} , \\
v_{i,k+1} &= v_{i,k} + \Delta T a_{i,k} , \\
\end{split}
\end{equation}
where $\Delta T \in \RRplus$ is the sampling time, $p_{i,k} \in \RR$ is the longitudinal position of the vehicle to the conflict point at time $k$, and  $v_{i,k} \in \RR$ and $a_{i,k} \in \RR$ are the speed and acceleration of the vehicle $i$ at time step $k$, respectively.
The vectors of states and control inputs of vehicle $i$ are defined by $\bb{x}_{i,k} = [p_{i,k}, v_{i,k}]^{\top}$ and $u_{i,k} = a_{i,k}$, respectively.
We consider the following state and input constraints
\begin{equation}
\label{eq:bound}
v_{\text{min}} \le v_{1,k+1} \le v_{\text{max}},\; u_{\text{min}} \le a_{i,k} \le u_{\text{max}}, \; \forall k \in \III_t,
\end{equation}
where $u_{\text{min}}$, $u_{\text{max}} \in \RR$ are the minimum deceleration and maximum acceleration, respectively, and $v_{\text{min}}$, $v_{\text{max}} \in \RR$ are the minimum and maximum speed limits, respectively.
Moreover, we impose the following safety constraint
\begin{equation}
\label{eq:safe-dist}
r^2 -  (p_{1,k+1}^2 + p_{2,k+1}^2) \le 0, \; \forall k \in \III_t,
\end{equation}
to guarantee that the predicted distances are greater than a safe distance, where $r \in \RRplus$ is a safety threshold.
The MPC objective is formed based on the idea of finding a Nash equilibrium of a potential game that models the interaction between the CAV and the HDV \cite{Le2022CDC} as given by
\begin{equation}
\label{eq:mpc-obj-func}
\sum_{k \in \III_t} \Big( \sum_{i = 1,2} l_{i} (\bb{x}_{i,k+1}, u_{i,k}) 
+ l_{12} (\bb{x}_{1,k+1}, \bb{x}_{2,k+1}) \Big),
\end{equation}
where $\III_t = \{ t, \dots, t+H-1 \}$ is the set of time steps in the control horizon of length $H \in \NN \setminus \{0\}$ at time step $t$.
The individual objective $l_i(\cdot),\, i = 1,2$ in \eqref{eq:mpc-obj-func} includes minimizing the control input for smoother movement and energy saving 
and minimizing the deviation from the maximum speed to reduce the travel time, 
\ie
\begin{equation}
\label{eq:ex-ego1}
l_{i} (\bb{x}_{i,k+1}, u_{i,k})
= \begin{bmatrix}
\omega_{i,1} \\
\omega_{i,2}
\end{bmatrix}^\top
\begin{bmatrix}
a_{i,k}^2 \\
(v_{i,k+1} - v_{i,\text{ref}})^2
\end{bmatrix} ,
\end{equation}
for $i = 1,2$, where $\omega_{i,1}, \omega_{i,2} \in \RRplus$ is the vector of positive weights and let denote $\bbsym{\omega}_i = [\omega_{i,1}, \omega_{i,2}]^\top$,
while $v_{i,\text{ref}}$ is the desired speed of \vehicle{i}. 
We consider that $v_{i,\text{ref}}$ can either be the maximum allowed speed $v_{\max}$ or the output of a car-following model if there is a preceding vehicle.
The shared objective function takes the form of a logarithmic penalty function of the distance between two vehicles as follows
\begin{equation}
\label{eq:ex-cooperative}
l_{12} (\bb{x}_{1,k+1}, \bb{x}_{2,k+1})
= - \omega_{12} \log \big( p_{1,k+1}^2 + p_{2,k+1}^2 + \epsilon \big) ,
\end{equation}
where $\omega_{12} \in \RRplus$ is a positive weight and $\epsilon \in \RRplus$ is a small positive number added to guarantee that the argument of the logarithmic function is always positive.

The MPC problem for \CAV{1} in this example is thus formulated as follows
\begin{equation}
\label{eq:ex-mpc}
\begin{split}
&
\underset{ \{u_{1,k}, u_{2,k}\}_{k \in \III_t} }{\minimize} \;  \quad \eqref{eq:mpc-obj-func} \\
& \text{subject to:}  \\
& \quad \text{\eqref{eq:integrator}}, \text{\eqref{eq:bound}}, \text{\eqref{eq:safe-dist}}, \; \forall k \in \III_t, \, i = 1,2.
\end{split}
\end{equation}
In the objective function of the MPC problem \eqref{eq:ex-mpc}, $\bbsym{\omega}_{2}$ and ${\omega}_{12}$ that best describe the human driving behavior can be learned online using moving horizon IRL.
For further details on the MPC formulation and moving horizon IRL implementation, the readers are referred to \cite{Le2023ACC} and \cite{Le2022CDC}, respectively.
Given the learned values of $\bbsym{\omega}_{2}$ and ${\omega}_{12}$,
the CAV's objective weights $\bbsym{\omega}_{1}$ can be adapted to achieve the desired performance.
In this example, the context and variable of contextual BO are $\bbsym{\theta} := \log_{10} \bbsym{\omega}_2$ and $\bbsym{z} := \log_{10} \bbsym{\omega}_1$, respectively.
The solution model learns the latent mapping from $\log_{10} \bbsym{\omega}_2$ to $\log_{10} \bbsym{\omega}_1^*$, \ie $\log_{10} \bbsym{\omega}_1^* \sim \GPModel_s(\log_{10} \bbsym{\omega}_2)$.
We fix the shared objective weight $\omega_{12} = 10^0$ as the solution of the MPC problem does not change if all the weights are scaled by a positive factor.
In addition, we consider the domain sets $\WWW_i = \{ \omega_{i,1}, \omega_{i,2} \; | \; 10^{-2} \le \omega_{i,1}, \omega_{i,2} \le 10^2 \}$, for $i = 1,2$.

Given the vehicle trajectories obtained from simulation, where we use MPC with a vector of the weights ${\bbsym{\omega} = [\bbsym{\omega}_1^\top, \bbsym{\omega}_2^\top,{\omega}_{12}]^\top}$, we define a \emph{time-energy efficiency with collision penalty} metric which is formed as follows
\begin{equation}
\label{eq:ex-true-cost}
\begin{multlined}
\tilde{J}_{\bbsym{\omega}} (\bb{x}_{\text{MPC}}, \bb{u}_{\text{MPC}}) = \lambda^{\mathrm{time}} t_{1,f} + \lambda^{\mathrm{acce}} \!\! \int_{t_0}^{t_{1,f}} u_1^2 (t)\; dt \\
+ \lambda^{\mathrm{coll}} \mathrm{sigmoid} \big( g^{\mathrm{coll}}(\bb{x}_{\text{MPC}}) \big), 
\end{multlined}
\end{equation}
where $\lambda^{\mathrm{time}}$, $\lambda^{\mathrm{acce}}$, and $\lambda^{\mathrm{coll}} \in \RRplus$ are constants. 
In \eqref{eq:ex-true-cost}, $t_{1,f}$ is the time that \CAV{1} exits the control zone, $\int_{t_0}^{t_{1,f}} u_1^2 (t)\; dt$ is the cumulative acceleration of \CAV{1} from $t_0 = 0$ to $t_{1,f}$,
while $\mathrm{sigmoid} \big( g^{\mathrm{coll}}(\bb{x}_{\text{MPC}}) \big)$ is the sigmoid penalty function to continuously approximate the indicator function of the safety constraint $g^{\mathrm{coll}}(\bb{x}_{\text{MPC}}) \le 0$ \cite{Le2023ACC}.
The function $g^{\mathrm{coll}}(\bb{x}_{\text{MPC}})$ is defined as the maximum of the left-hand side in \eqref{eq:safe-dist} for the entire trajectory.
We consider the performance metric in contextual BO as the negative average of $\tilde{J}_{\bbsym{\omega}} (\bb{x}_{\text{MPC}}, \bb{u}_{\text{MPC}})$ across multiple simulations with $n_s \in \NN$ i.i.d. initial positions and speeds of the vehicles, \ie
\begin{equation}
J (\bbsym{z}, \bbsym{\theta}) := - \frac{1}{n_s} \sum_{n = 1}^{n_s} \tilde{J}_{\bbsym{\omega}} \Big( \bb{x}^{(n)}_{\text{MPC}}, \bb{u}_{\text{MPC}}^{(n)} \Big),
\end{equation}
where $\Big( \bb{x}^{(n)}_{\text{MPC}}, \bb{u}_{\text{MPC}}^{(n)} \Big)$ denotes the state and input trajectories in the $n$-th simulation.

\begin{figure*}[!tb]
\centering
\begin{subfigure}{\textwidth}
\centering
\begin{subfigure}{.4\textwidth}
\centering
\includegraphics[scale=0.24]{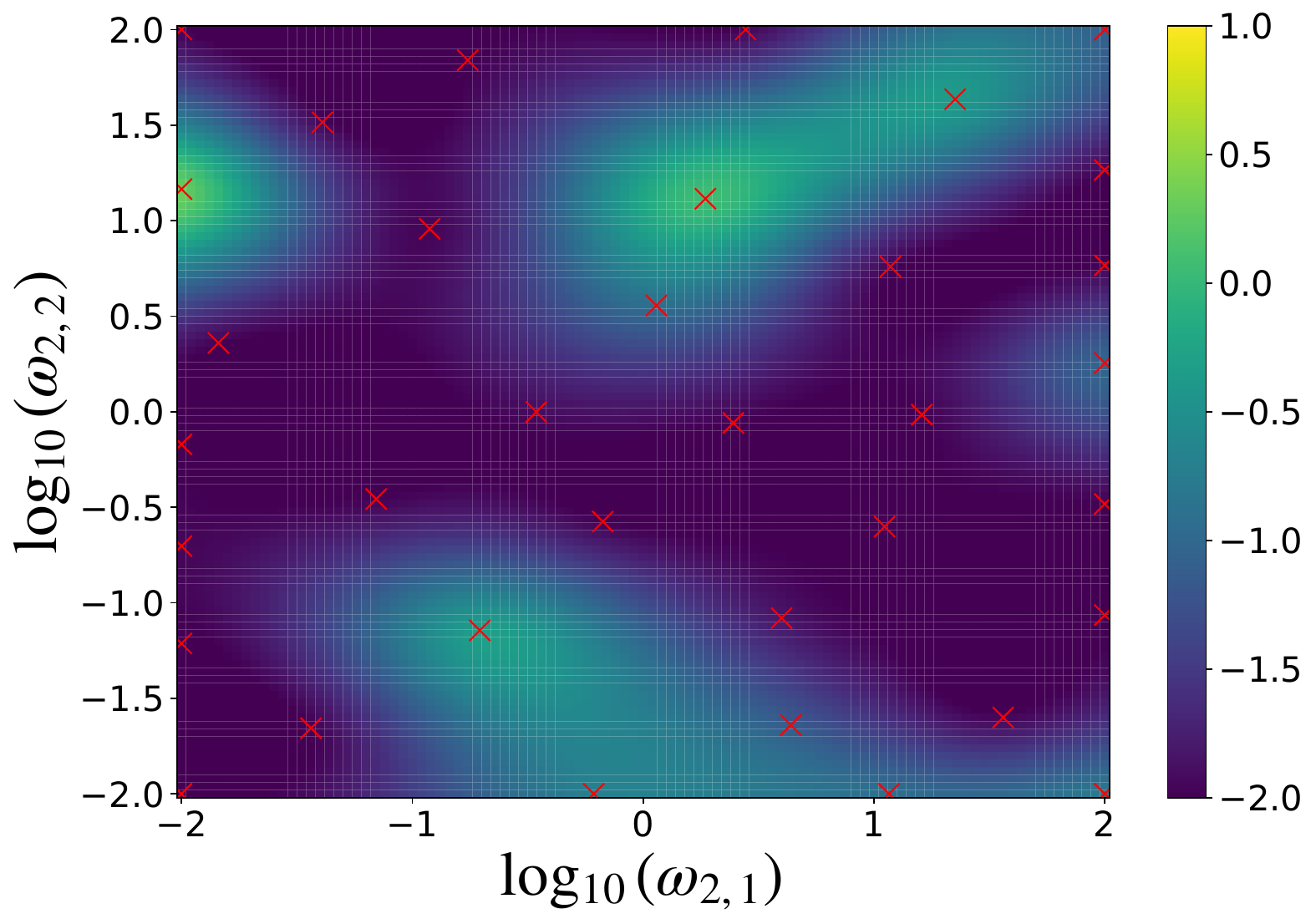}
\end{subfigure}
\begin{subfigure}{.4\textwidth}
\centering
\includegraphics[scale=0.24]{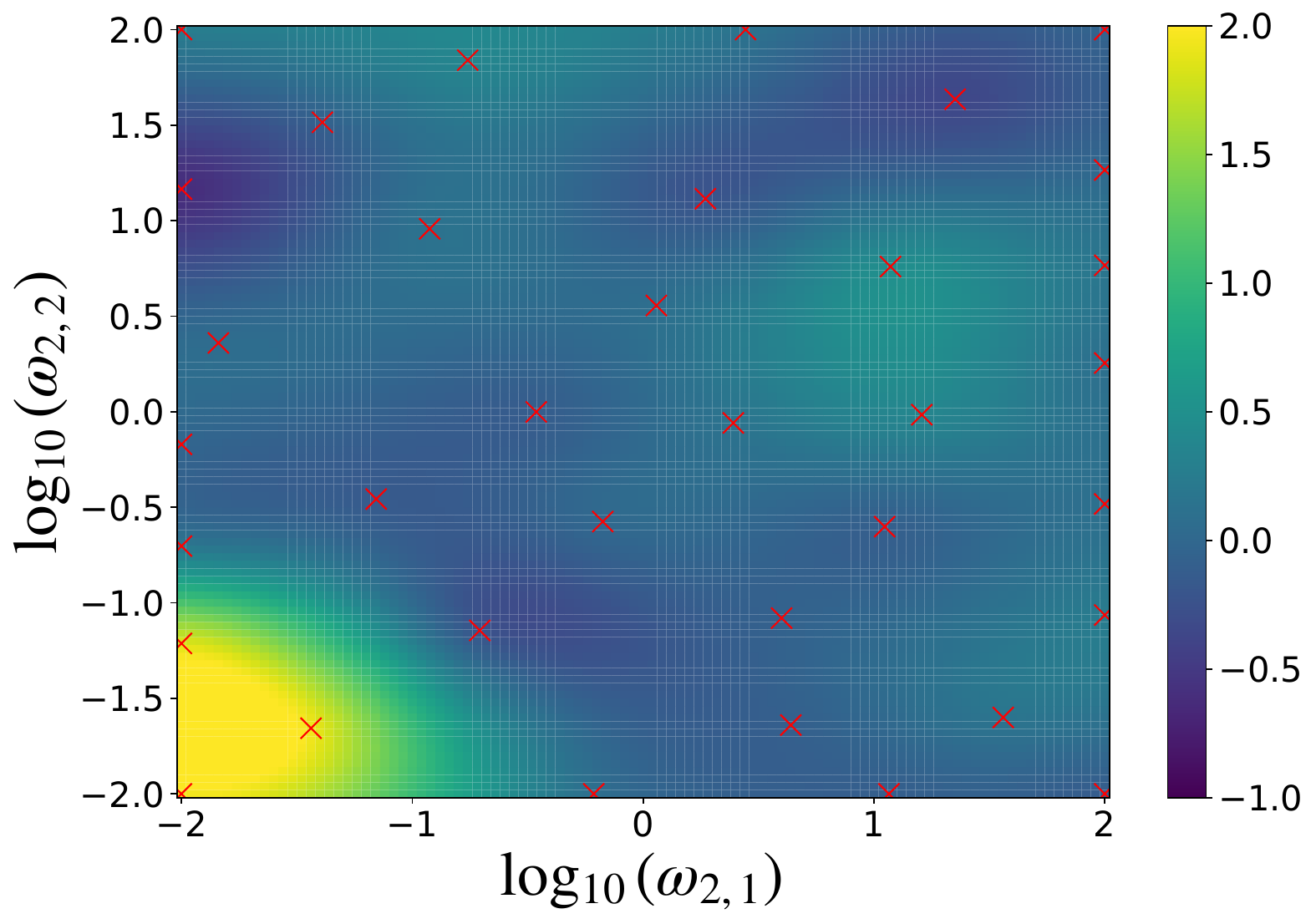}
\end{subfigure}
\caption{Heat maps for $\log_{10} (\omega_{1,1})$ (left) and $\log_{10} (\omega_{1,2})$ (right) using the Mat\'ern $3/2$ kernel.}
\end{subfigure}
\begin{subfigure}{\textwidth}
\centering
\begin{subfigure}{.4\textwidth}
\centering
\includegraphics[scale=0.24]{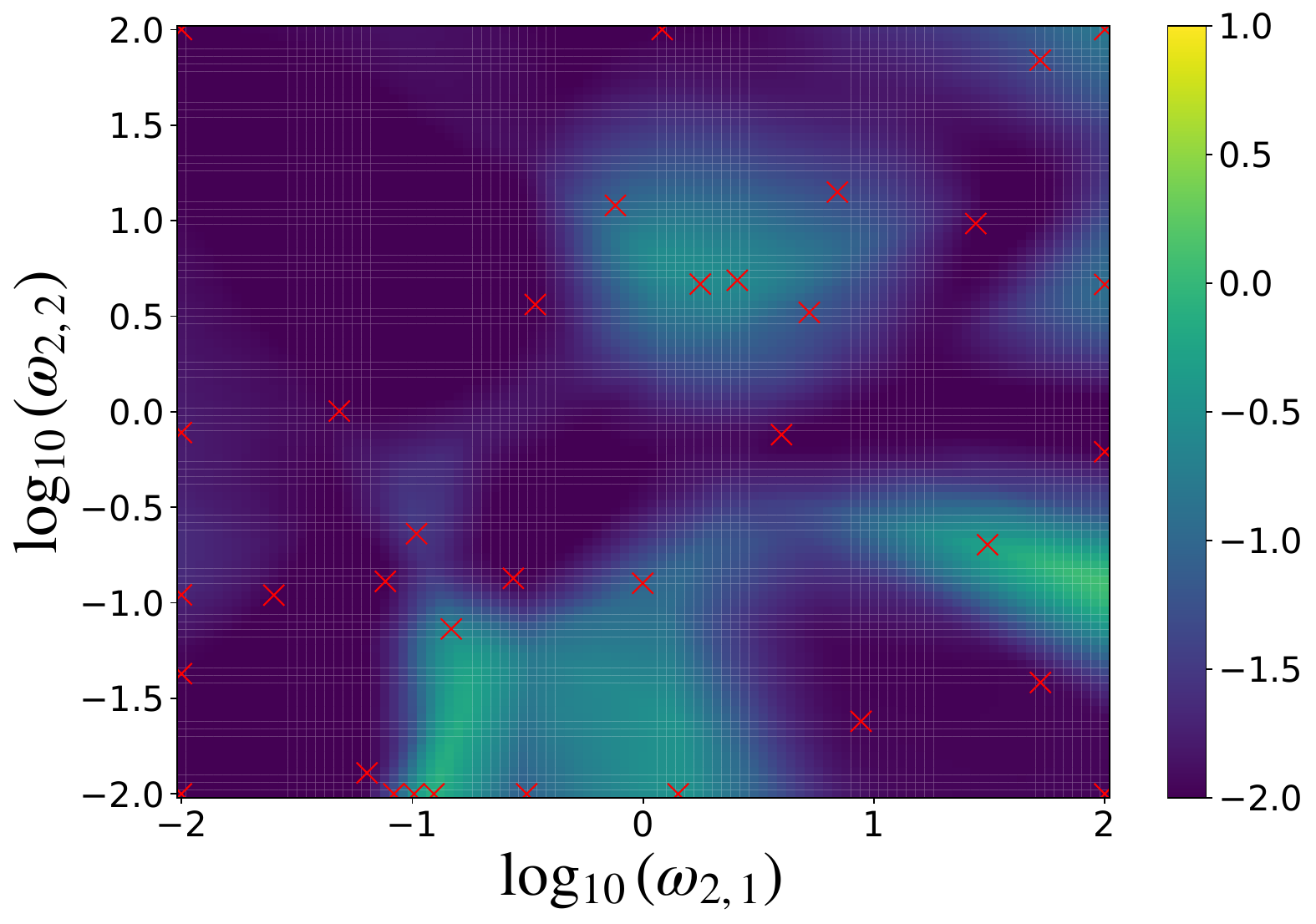}
\end{subfigure}
\begin{subfigure}{.4\textwidth}
\centering
\includegraphics[scale=0.24]{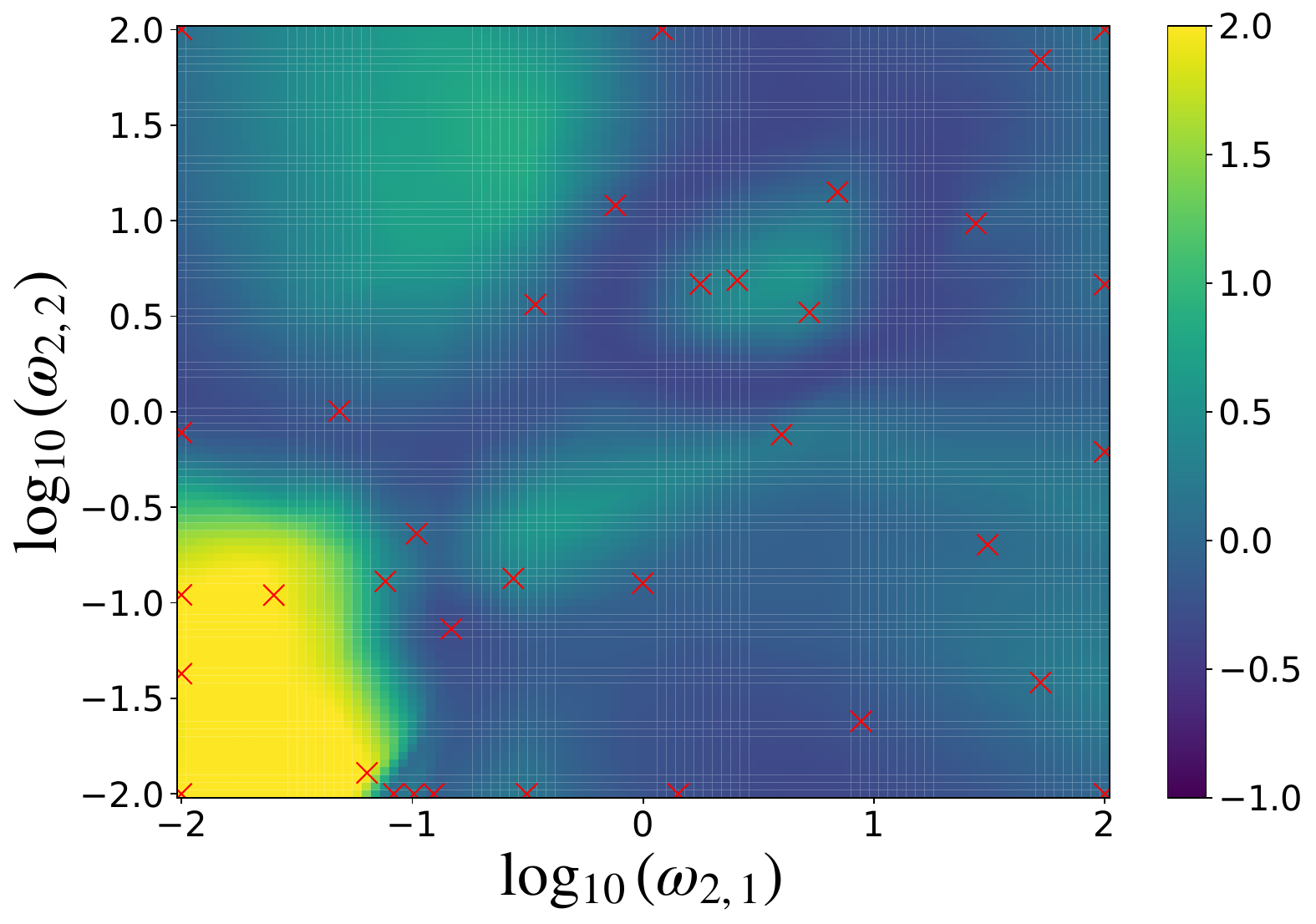}
\end{subfigure}
\caption{Heat maps for $\log_{10} (\omega_{1,1})$ (left) and $\log_{10} (\omega_{1,2})$ (right) using the deep kernel learning.}
\end{subfigure}
\caption{Comparison of optimal weight adaptation strategies using different kernels. The red crosses represent sampled values of $\log_{10} (\bbsym{\omega}_{2})$.}
\label{fig:colormap}
\end{figure*}

In our implementation, we choose the following parameters: 
$\lambda^{\mathrm{time}} = 1.0$, 
$\lambda^{\mathrm{acce}} = 5.0$, 
$\lambda^{\mathrm{coll}} = 10^4$, 
$k_{\mathrm{max}} = 30$, 
$j_{\mathrm{max}} = 30$, 
and $\beta = 10^2$.
We compare the weight adaptation strategies obtained by using two different kernels for learning the solution model: (1) the Mat\'ern $3/2$ kernel and (2) the deep learning kernel.
The deep learning kernel is constructed using a radial basis function (RBF) base kernel and a deep neural network with three hidden layers and 64 neurons per layer.
In both cases, the Mat\'ern $3/2$ kernel is used for the surrogate objective model.
In Fig.~\ref{fig:colormap}, we illustrate the learned weight adaptation strategies obtained from the proposed framework using the two kernels, shown as heat maps.
The sampling locations for the contexts are indicated by red crosses.
We observe that the adaptive sampling algorithm distributes the context samples across the domain, with more samples in regions where the output changes sharply or near the domain boundaries to better capture the solution model.
Overall, although the deep learning kernel is effective at capturing sharp variations in the latent model, it may be more sensitive to potentially inexact inner-loop solution data.
In contrast, the Mat\'ern $3/2$ kernel results in a smoother adaptation strategy.
Thus, in the simulations and experiments presented next, we utilize the GP solution model with the Mat\'ern $3/2$ kernel.

\subsection{Comparison using Simulation Results}

\begin{table*}[!bt]
\caption{Comparison between adaptive MPC using our proposed framework (\#1) and using the method in \cite[Appendix~D]{char2019offline} (\#2), along with three non-adaptive MPC designs.}
\label{tab:compare} 
\centering
\begin{tabular}{ K{0.3\textwidth} | K{0.12\textwidth} K{0.12\textwidth} K{0.1\textwidth} K{0.1\textwidth} K{0.1\textwidth}}
\toprule[1pt]
\backslashbox{Comparison metrics}{Controllers} & {Adaptive MPC \#1 (our approach)} & {Adaptive MPC \#2} & {Non-adaptive MPC \#1} & {Non-adaptive MPC \#2} & {Non-adaptive MPC \#3} \\
\midrule[0.5pt]
Number of simulations with safety & $9997$ & $9970$ & $10000$ & $9796$ & $9989$ \\
Average travel time ($s$) & $11.6$ & $12.9$ & $15.8$ & $11.7$ & $17.5$ \\
Average acceleration ($m/s^2$) & $0.130$ & $0.127$ & $0.119$ & $0.134$ & $0.098$ \\
\bottomrule[1pt]
\end{tabular}
\end{table*}

Before applying the proposed framework in experiments, we evaluate its benefits by comparing the performance of the MPC utilizing the adaptation strategy against three distinct fixed-weight MPC designs across a significant number of testing simulations.
The values of $\bbsym{\omega}_1$ in three non-adaptive MPC designs are chosen as $[10^{-1.0}, 10^{0.0}]^\top$, $[10^{-1.0}, 10^{0.5}]^\top$, and $[10^{0.25}, 10^{-0.25}]^\top$, which are manually tuned to prioritize safety, time efficiency, and acceleration efficiency, respectively.
Moreover, we compare our proposed framework with the weight adaptation strategy obtained using the method in \cite[Appendix~D]{char2019offline}, in which the approximate solutions for new contexts are found by optimizing the posterior mean of the surrogate model.
Note that our implementation involves a modification to \cite[Appendix~D]{char2019offline}, where adaptive sampling using conditional entropy maximization is utilized rather than Thompson sampling.
We compare the performance of the controllers based on three metrics: (1) the number of simulations with safety, (2) average travel time, and (3) average acceleration.
Those comparison metrics are computed by averaging $10,000$ randomized simulations with different initial positions and speeds of the vehicles and heterogeneous driving styles of the human drivers.
The human driving actions in the simulations are generated using an IRL model \cite{kuderer2015learning}, where the weights are randomized to emulate various driving behaviors.

The metrics collected for all controllers can be found in Table~\ref{tab:compare}.
The results reveal that while maintaining a similar level of safety, adaptive MPC \#1 (our framework) significantly outperforms non-adaptive MPC \#1 and \#3 in terms of travel time. 
However, adaptive MPC \#1 requires a higher acceleration rate compared to non-adaptive MPC \#1 and \#3. 
On the other hand, given the same level of time and acceleration efficiency, adaptive MPC \#1 demonstrates a higher safety level than non-adaptive MPC \#2.
Hence, the comparison implies that employing the learned weight adaptation strategy enables adaptive MPC to achieve a delicate balance between conservativeness and aggressiveness in designing the MPC.
In comparison with adaptive MPC \#2, our proposed framework demonstrates slightly better performance in terms of safety and average travel time, with similar acceleration efficiency.

Finally, the main advantage of our framework compared to the related method in \cite[Appendix~D]{char2019offline} is that our framework does not require solving a complex optimization problem to find approximate solutions for new contexts, which makes it more suitable for real-time controller adaptation.
From our simulations, the average computation time for GP predictions is \SI{2.4}{ms}, while that for IRL and solving the MPC problem is \SI{17.8}{ms}.
On the other hand, when we tried solving $\bbsym{z}^* = \argmax \mu_o (\bbsym{z}, \bbsym{\theta}^{(k)})$ for a GP model with $100$ data points, using the particle swarm optimization algorithm, it took at least \SI{400}{ms}.
Moreover, the experiments we conducted (see Section~\ref{subsec:exp}) verify that the proposed framework can be used in real time.

\subsection{Experimental Validation}
\label{subsec:exp}

\begin{figure*} 
\centering
\begin{subfigure}{0.24\textwidth} \hspace{-10pt}
\centering
\includegraphics[width=1.0\textwidth]{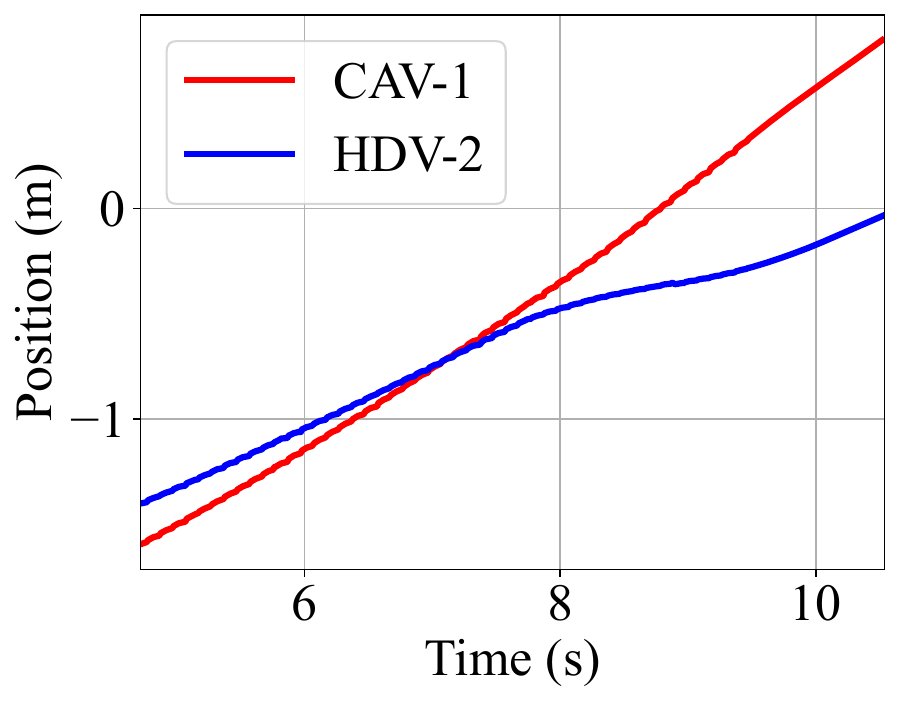}
\caption{Experiment $\#1$}
\end{subfigure} 
\begin{subfigure}{0.24\textwidth} \hspace{-10pt}
\centering
\includegraphics[width=1.0\textwidth]{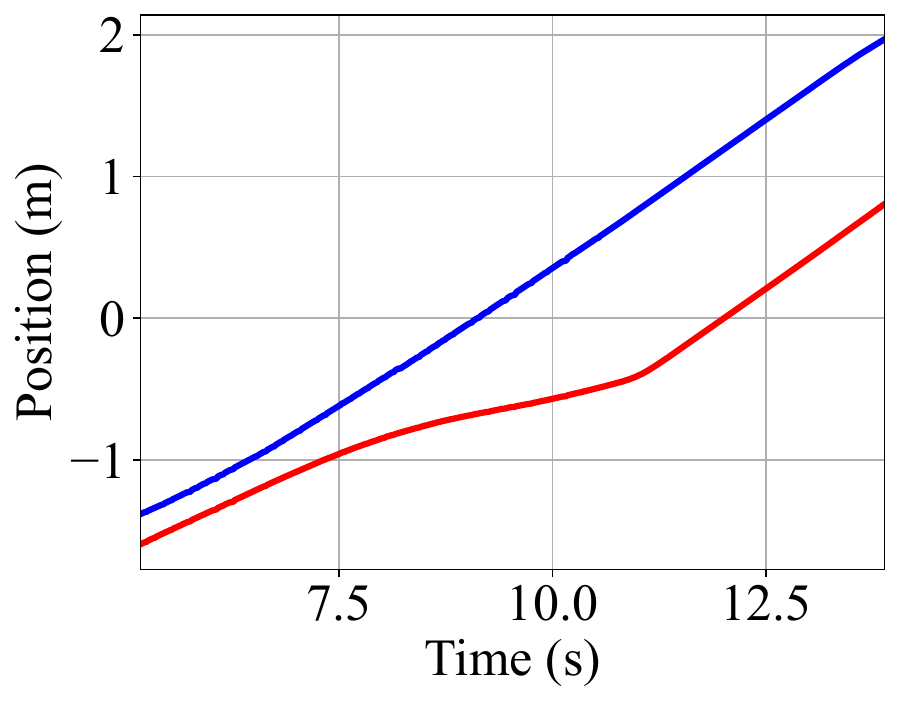}
\caption{Experiment $\#2$}
\end{subfigure} 
\begin{subfigure}{0.24\textwidth} \hspace{-10pt}
\centering
\includegraphics[width=1.0\textwidth]{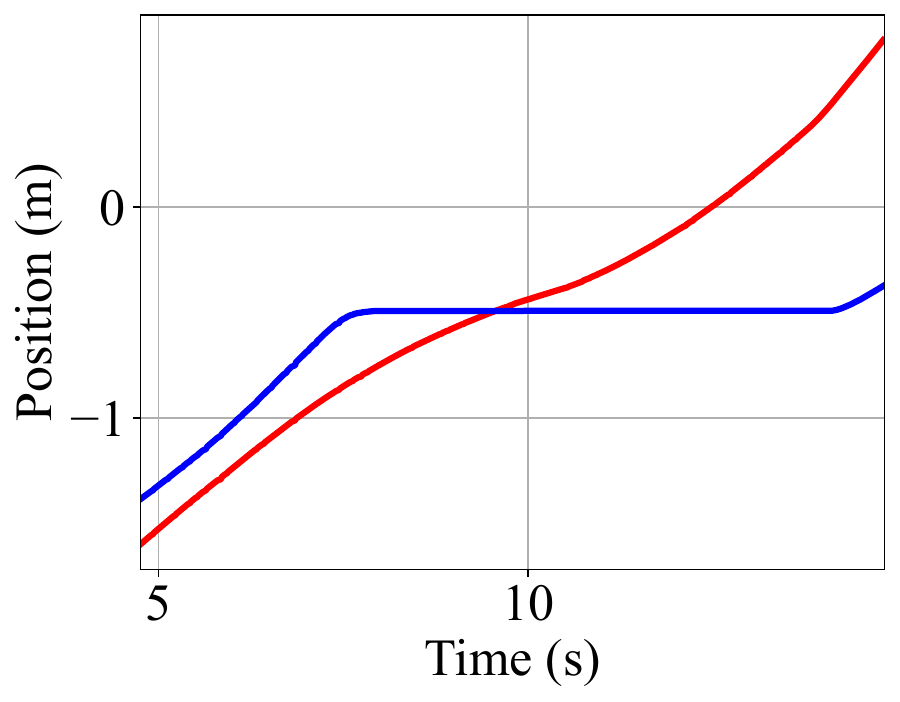}
\caption{Experiment $\#3$}
\end{subfigure} 
\begin{subfigure}{0.24\textwidth} \hspace{-10pt}
\centering
\includegraphics[width=1.0\textwidth]{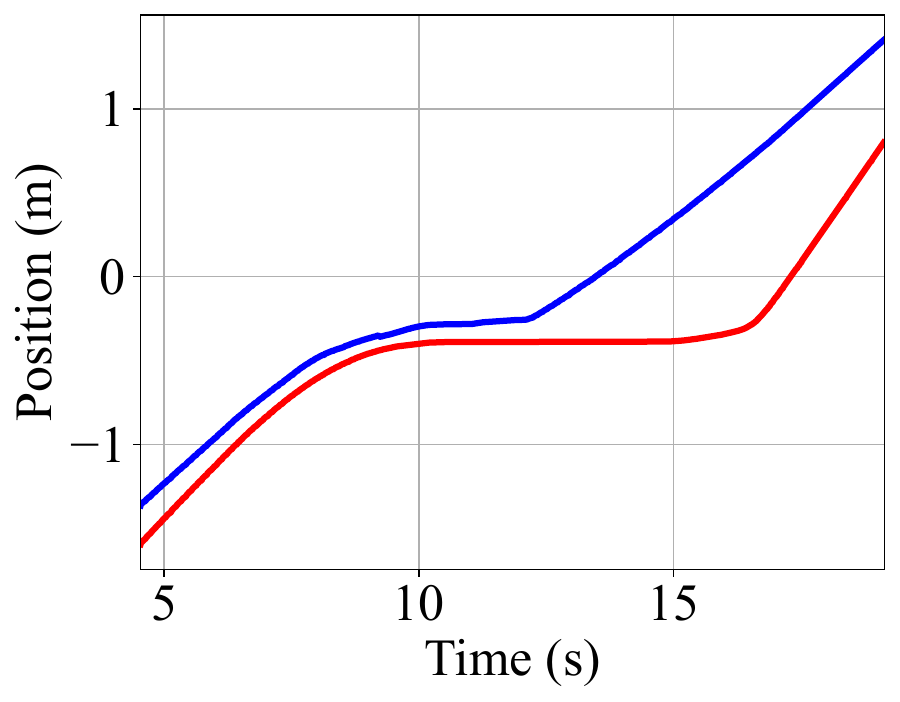}
\caption{Experiment $\#4$}
\end{subfigure} 

\begin{subfigure}{0.24\textwidth} \hspace{-10pt}
\centering
\includegraphics[width=1.0\textwidth]{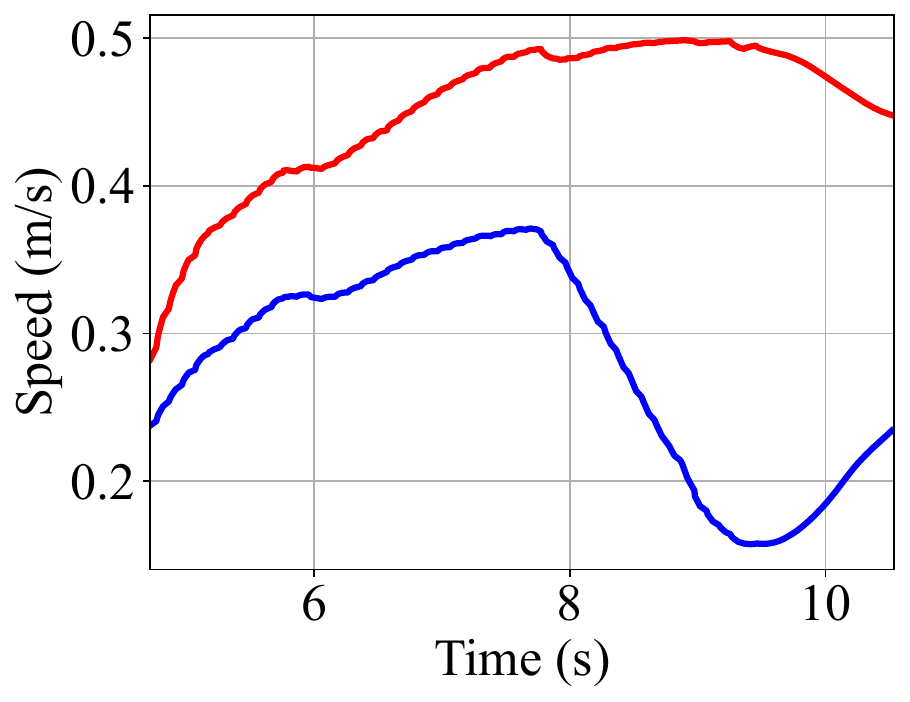}
\caption{Experiment $\#1$}
\end{subfigure} 
\begin{subfigure}{0.24\textwidth} \hspace{-10pt}
\centering
\includegraphics[width=1.0\textwidth]{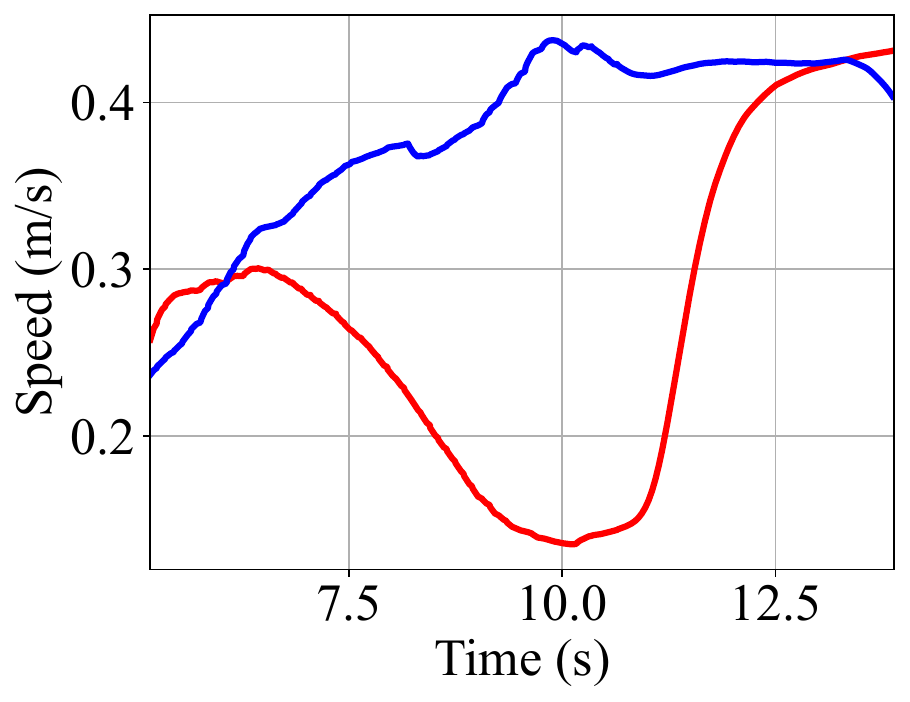}
\caption{Experiment $\#2$}
\end{subfigure} 
\begin{subfigure}{0.24\textwidth} \hspace{-10pt}
\centering
\includegraphics[width=1.0\textwidth]{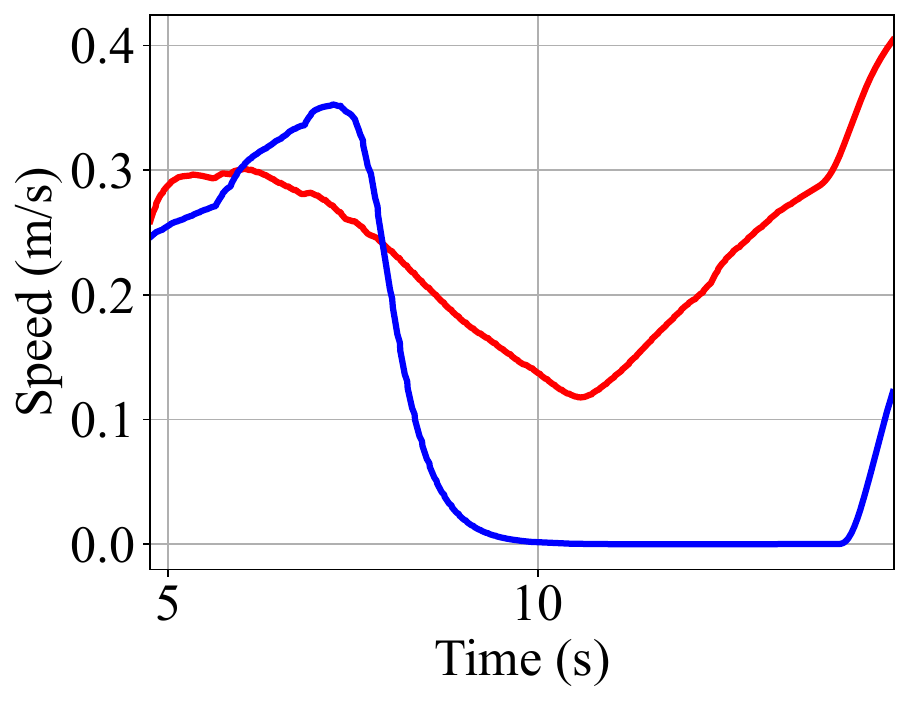}
\caption{Experiment $\#3$}
\end{subfigure} 
\begin{subfigure}{0.24\textwidth} \hspace{-10pt}
\centering
\includegraphics[width=1.0\textwidth]{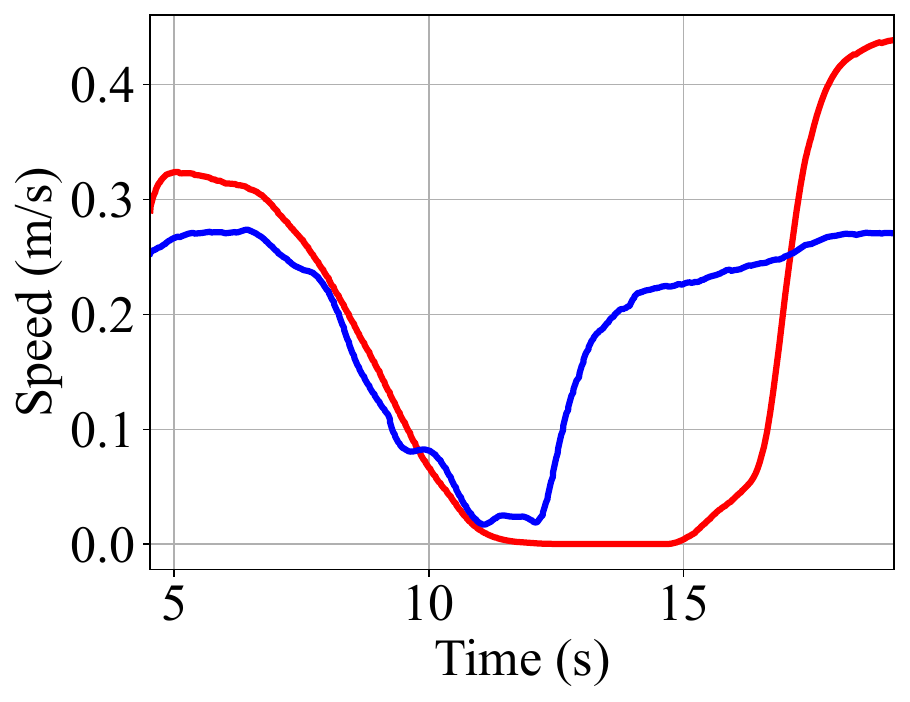}
\caption{Experiment $\#4$}
\end{subfigure} 
\caption{Position trajectories (top figures) and speed profiles (bottom figures) of vehicles in the experiments involving one HDV.}
\vspace{-5mm}
\label{fig:traj}
\end{figure*}

We validate the proposed framework through experiments in the \texttt{IDS3C} testbed \cite{chalaki2021CSM}.
The \texttt{IDS3C} has $1:24$ scaled robotic cars that can function as either CAVs or HDVs.
Each robotic car is equipped with a Raspberry Pi embedded computer for wireless communication and local computation, such as low-level lane-tracking control.
For the HDV option, we use Logitech G29 driving emulators to allow human participants to manually control the robotic cars.
The participants can observe the environment through a camera mounted on each robotic car, and make real-time driving decisions by adjusting speed and steering via the driving emulators.
In addition to the fully manual driving mode, we integrated a driving mode with an advanced driver-assistance system, where human drivers control only the vehicle's speed while a lateral control algorithm manages the steering.
This setup enables realistic human driving behavior for experimental purposes.
Real-time localization of the robotic cars is provided by a VICON motion capture system. 
A central mainframe computer (equipped with an Intel® Xeon® w9-3475X CPU) handles data acquisition from the VICON system and the driving emulators, performing computations for the proposed framework to obtain the control commands for the vehicles. 
The control commands are transmitted from the mainframe computer to the Raspberry Pi embedded onboard each robotic car via the UDP/IP protocol.
Videos of the experiments can be found at \url{https://sites.google.com/cornell.edu/mt-mpc-exp}.

\begin{figure*} 
\centering
\begin{subfigure}{0.24\textwidth} \hspace{-10pt}
\centering
\includegraphics[width=1.0\textwidth]{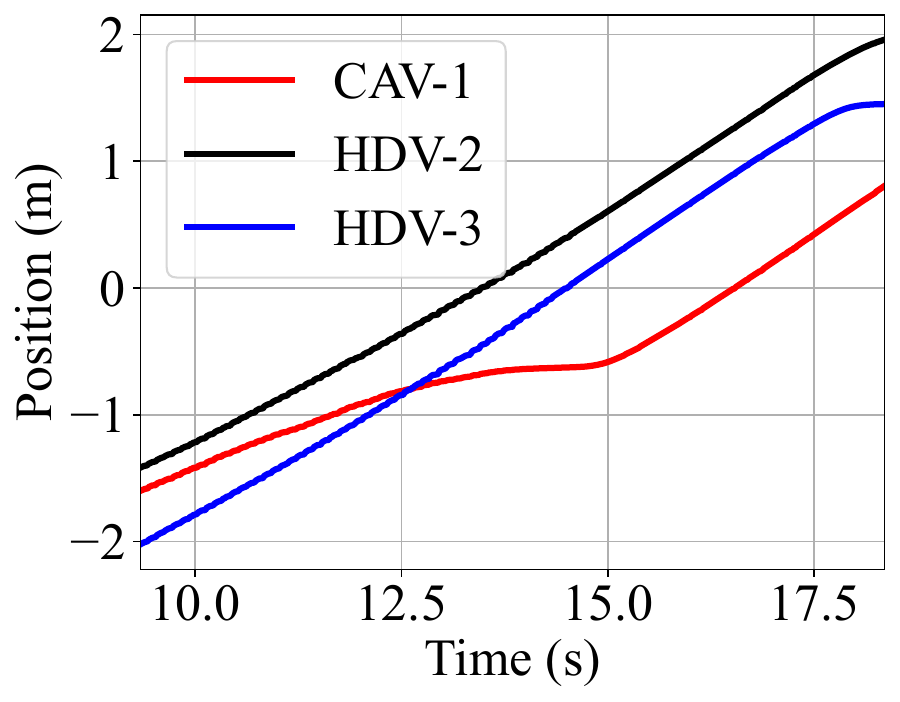}
\caption{Experiment $\#1$}
\end{subfigure} 
\begin{subfigure}{0.24\textwidth} \hspace{-10pt}
\centering
\includegraphics[width=1.0\textwidth]{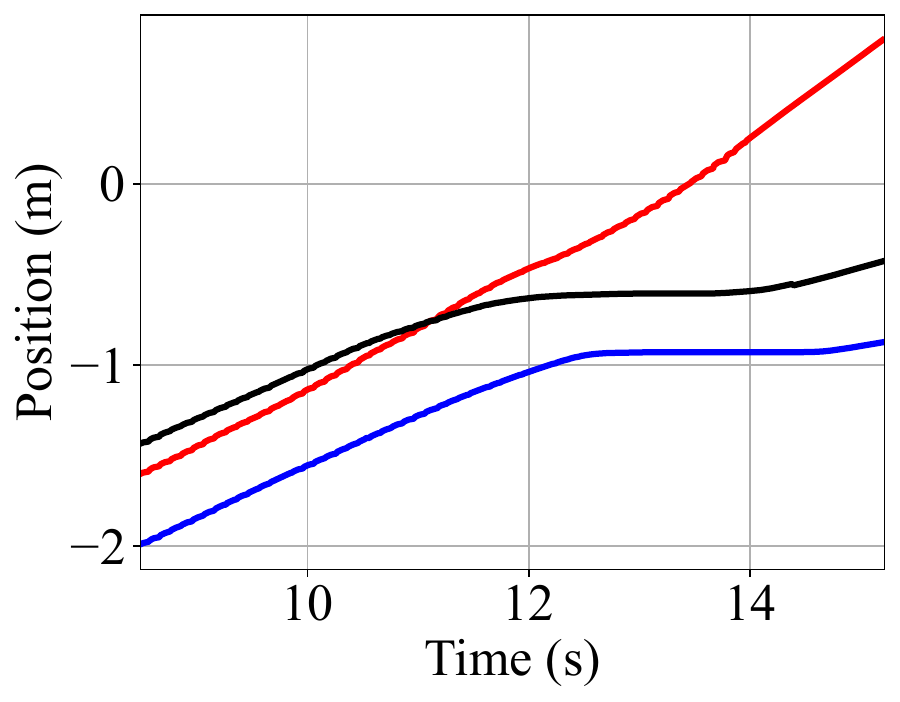}
\caption{Experiment $\#2$}
\end{subfigure} 
\begin{subfigure}{0.24\textwidth} \hspace{-10pt}
\centering
\includegraphics[width=1.0\textwidth]{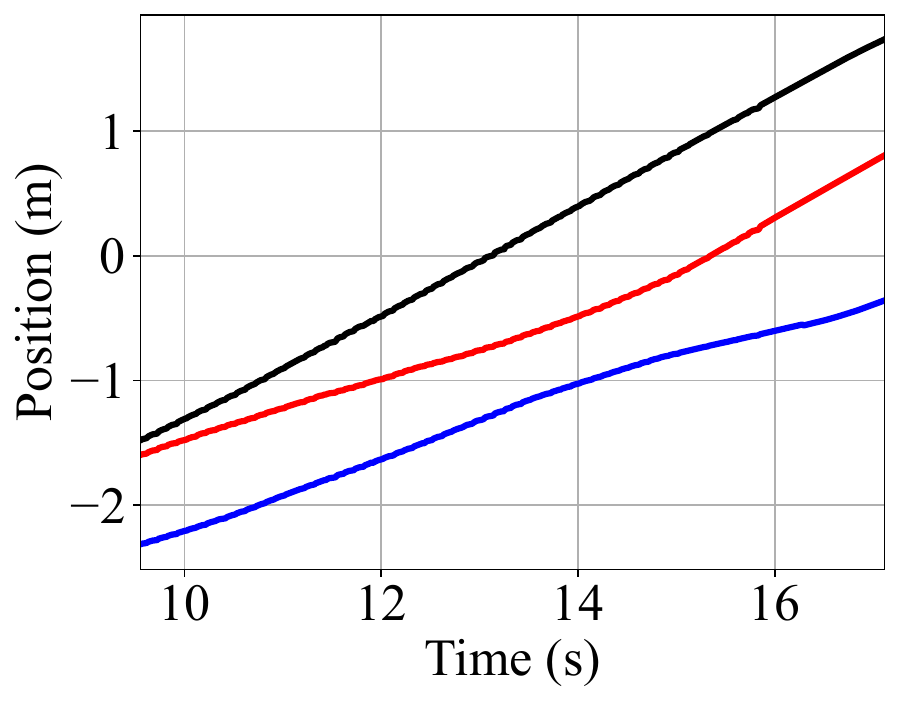}
\caption{Experiment $\#3$}
\end{subfigure} 

\begin{subfigure}{0.24\textwidth} \hspace{-10pt}
\centering
\includegraphics[width=1.0\textwidth]{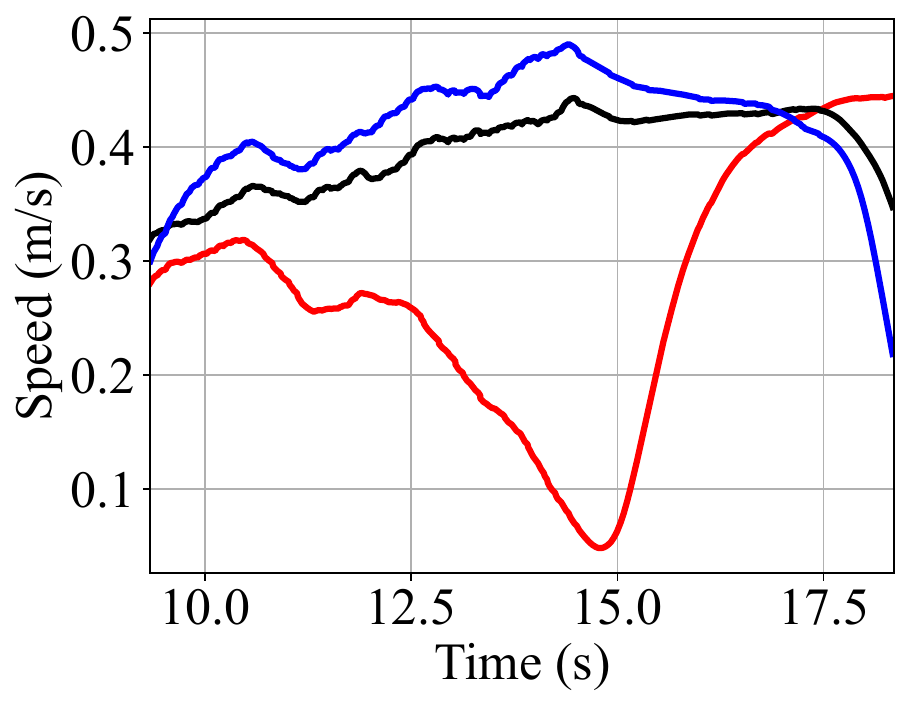}
\caption{Experiment $\#1$}
\end{subfigure} 
\begin{subfigure}{0.24\textwidth} \hspace{-10pt}
\centering
\includegraphics[width=1.0\textwidth]{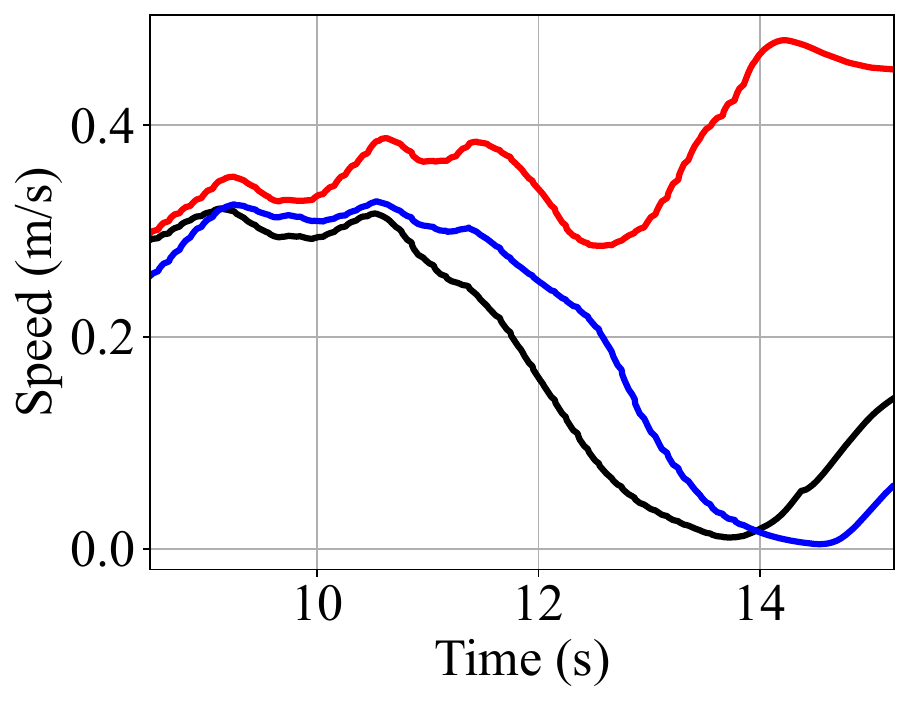}
\caption{Experiment $\#2$}
\end{subfigure} 
\begin{subfigure}{0.24\textwidth} \hspace{-10pt}
\centering
\includegraphics[width=1.0\textwidth]{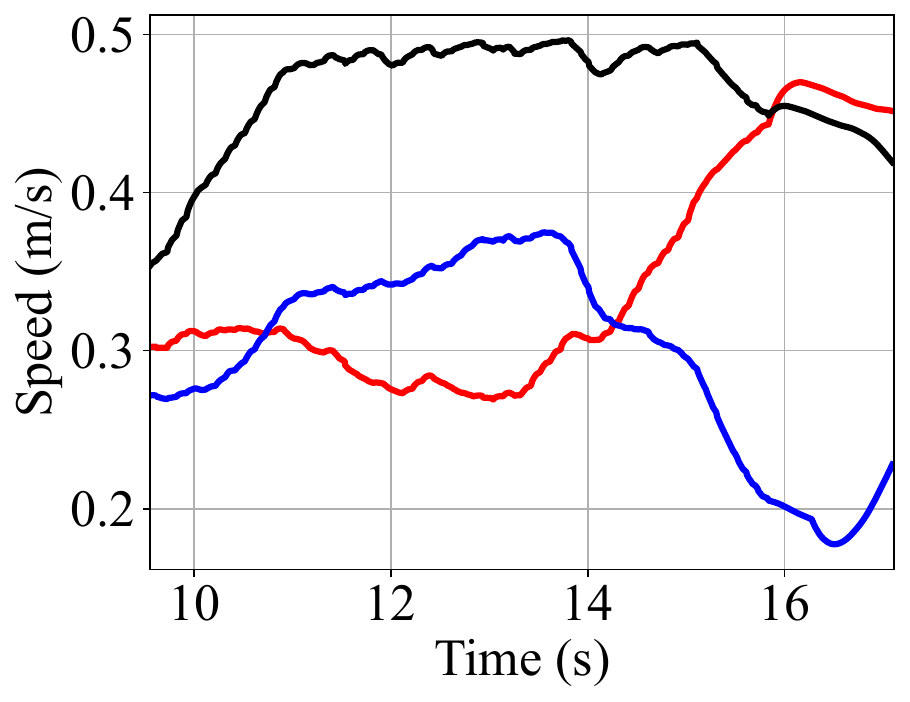}
\caption{Experiment $\#3$}
\end{subfigure} 
\caption{Position trajectories (top figures) and speed profiles (bottom figures) of vehicles in the experiments involving two HDVs.}
\vspace{-5mm}
\label{fig:traj_2hdv}
\end{figure*}

In Fig.~\ref{fig:traj}, we show the position trajectories and speed profiles of the two vehicles in four specific simulations, each demonstrating different driving styles generated by the human participant.
As illustrated in the figure, the MPC with learned weight adaptation effectively adjusts the behavior of CAVs to accommodate diverse human driving styles, resulting in safe interactions between vehicles. 
For example, when the HDV exhibits an aggressive driving style, the CAV compensates by behaving more conservatively. 
Conversely, if the HDV is more cautious, the CAV may adopt a more aggressive behavior.
We also validate the controller in a more challenging scenario where a CAV attempts to cross an intersection with two HDVs traveling in the conflicting lane relative to the CAV.
In this scenario, although two HDVs are present in the experiments, the MPC considers only one HDV at a time, with its weights adapted based on the IRL model for that HDV, so that we can exploit the MPC formulation presented in Section~\ref{subsec:mpc}.
Specifically, the MPC considers the first HDV that has not yet crossed the conflict point.
The position trajectories and speed profiles of the vehicles in three specific simulations, where the crossing orders vary depending on the behavior of the HDVs, are shown in Fig.~\ref{fig:traj_2hdv}.
The results confirm that the MPC obtained using our framework is effective even in more challenging scenarios involving multiple human drivers.

\section{Conclusions} 
\label{sec:concl}

In this letter, we proposed a framework to address the controller adaptation problem for dynamic systems with task-dependent or time-varying parameters via learning the solutions of contextual BO with GPs. 
We demonstrated the efficacy of the framework through a sim-to-real application, where the weighting strategy of MPC for CAVs interacting with HDVs is learned from simulations and applied in real-time experiments. 
We also conducted massive simulations to demonstrate the advantages of the adaptive MPC against non-adaptive MPC designs.
The proposed framework can be enhanced by (1) incorporating black-box hard constraints for safety-critical applications and (2) improving scalability for high-dimensional problems.
These extensions will be considered in future work.

\bibliographystyle{IEEEtran}
\bibliography{IEEEabrv,references,references_IDS}

\balance

\end{document}